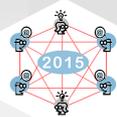

# Modeling Quantum Optical Components, Pulses and Fiber Channels Using OMNeT++

Ryan D. L. Engle*, Douglas D. Hodson*, Michael R. Grimaila*,
Logan O. Mailloux*, Colin V. McLaughlin† and Gerald Baumgartner‡
*Air Force Institute of Technology, Wright-Patterson AFB, OH, USA
email: {douglas.hodson, michael.grimaila}@afit.edu
†Naval Research Laboratory, Washington, D.C., USA
‡Laboratory for Telecommunication Sciences, College Park, MD, USA

*Abstract*—QKD is an innovative technology which exploits the laws of quantum mechanics to generate and distribute unconditionally secure cryptographic keys. While QKD offers the promise of unconditionally secure key distribution, real world systems are built from non-ideal components which necessitates the need to model and understand the impact these non-idealities have on system performance and security. OMNeT++ has been used as a basis to develop a simulation framework to support this endeavor. This framework, referred to as "qkdX," extends OMNeT++'s module and message abstractions to efficiently model optical components, optical pulses, operating protocols and processes. This paper presents the design of this framework including how OMNeT++'s abstractions have been utilized to model quantum optical components, optical pulses, fiber and free space channels. Furthermore, from our toolbox of created components, we present various notional and real QKD systems, which have been studied and analyzed.

## I. INTRODUCTION

Modeling and Simulation (M&S) is often used as an efficient means to understand complex systems and their dynamics, as the model developer (i.e., a researcher or tester) is forced to fully understand the behaviors of interest, their inputs, and expected outputs in order to accurately simulate the system behavior. By systematically defining and decomposing the complex behaviors of interest, one is able to construct representative models, with varying levels of abstraction and generalization, needed to answer research questions of interest.

In this paper, we present a model and simulation framework to study Quantum Key Distribution (QKD) systems. Quantum Key Distribution (QKD) systems offer the promise to generate and distribute unconditionally secure cryptographic keys [1]. However, real world QKD systems are built from non-ideal components which differ greatly from their ideal counterparts [2], [3]. Our research is focused upon understanding and quantifying the impact component non-idealities have on QKD system performance and security. To achieve our objectives, we have extended the OMNeT++ by developing a quantum key distribution eXperimentation (qkdX) framework to model electrical, optical, and electro-optical components necessary to efficiently simulate complete QKD systems. Using the developed framework, we have designed, constructed, and simulated various QKD systems in order to study non-idealities in real-world implementations [4].

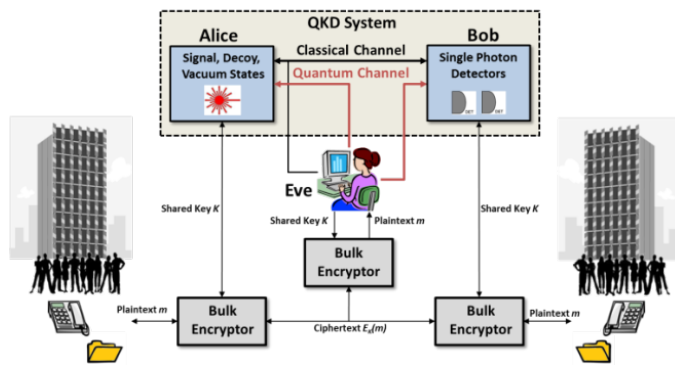

Fig. 1: QKD System Context Diagram

The qkdX framework itself consists of a set of modular components specific to the optics domain, so that QKD systems can be assembled. Using the abstractions for modules, messages and channels, components in the domain of optical systems, such as, beamsplitters, attenuators, optical pulses and fiber channels have been modeled.

This paper presents how we have leveraged OMNeT++ functionality and tailored abstractions to the domain of optical components. We first present a short background on QKD to provide a motivation for this research, followed sections on how the framework is organized, and how OMNeT++'s [5] abstractions have been utilized to model quantum optical components, optical pulses, fiber and free space channels. Finally, we conclude with three specific simulation studies supported by this framework: a polarization-based, prepare and measure BB84 QKD reference architecture, a decoy state enabled QKD architecture and a Measurement Device Independent (MDI) QKD architecture. The results of these studies demonstrate the value of using OMNeT++ as basis for complex system representation, simulation, and analysis.

## II. QKD BACKGROUND

The genesis of QKD can be traced back to Wiesner's idea of encoding messages on photons using polarized photons in conjugate bases to securely communicate information as quantum bits, called "qubits" [6], [7]. In 1984, Bennett and Brassard expanded this idea and proposed the first QKD protocol, known as BB84, to securely distribute cryptographic key between two parties, typically identified as Alice and Bob [1]. BB84 is a "prepare and measure" protocol where Alice prepares and transmits qubits to Bob who measures them. By exploiting the laws of quantum mechanics and performing statistical analysis of errors during "quantum transmission," QKD systems possess the inherent and unique capability to detect eavesdropping. Thus, theoretical QKD systems can deliver unconditionally secure keys to authenticated parties for use in cryptographic systems. A notional QKD system architecture is shown in Figure 1. An eavesdropper, Eve, is shown attached to the classical and quantum channels linking Alice and Bob. This situation illustrates the sort of contested environment in which QKD systems are designed to operate. An excellent introduction to QKD can be found here [8] and [2].

### III. Framework Packages & Organization

The primary objective of qkdX is to enable the rapid and efficient modelling and analysis of current and proposed QKD system implementations using varying levels of model abstraction [2]. From a software point of view, the qkdX framework is a library that defines specific OMNeT++-based modules for optical components (e.g., beamsplitters, attenuators, lasers, classical optical detectors, single photon detectors). The library also defines an abstract message class for which a few concrete pulse types have been defined. Finally, the framework defines a few specific channel types to model the propagation effects associated with optical pulses in fiber cables and free space. For example, an important aspect modeled with channels is the attenuation associated with propagating an optical pulse from one place to another (i.e., one module to another).

As shown in Figure 2, organizationally, OMNeT++ provides the fundamental infrastructure to build and interconnect system components and qkdX defines a set of abstract and concrete system models and components common to many different QKD architectures. End user simulation products crafted to support specific studies and analyzes are then constructed from these components and others (e.g., INET-based [9] components).

### IV. Optical Pulses

Much like the flow of classical communication packets, the flow or propagation of quantum optical pulses within the framework are carried by messages, specifically, a specialized message class that manages a pointer to an abstract pulse class as shown in Figure 3. Two concrete types of pulse representations are current defined, CoherentState [10] and FockState [11] - each provides features appropriate to model quantum effects associated with QKD systems.

The base Pulse class defines common attributes associated with all pulses, including wavelength, duration, a global phase which indicates a relative phase offset between it and a reference. Pulses also include polarization information and a characteristic shape. For a Fock state, the shape defines a probability amplitude which is used to determine when a photon arrives at a specific device - for a pulse of this type, the number of photons the pulse carries is known, thus its energy is known. For a coherent state representation, the number of photons associated with a pulse is probabilistic, so the shape serves a dual purpose; first to determine the amount of energy contained, and secondly, the time of arrival for zero or more photons. The shape of the pulse is integrated and combined with other parameters (e.g., such as e-field amplitutde) to determine its energy. This energy is divided by the energy of a photon at this wavelength to determine a mean photon number (i.e., MPN). This MPN is used as an input to a Poisson distribution to determine the number of photons present. The pulse shape is then normalized, used to construct a reverse cumulative distribution function, and a uniform random number is drawn to determine the arrival times of the photons.

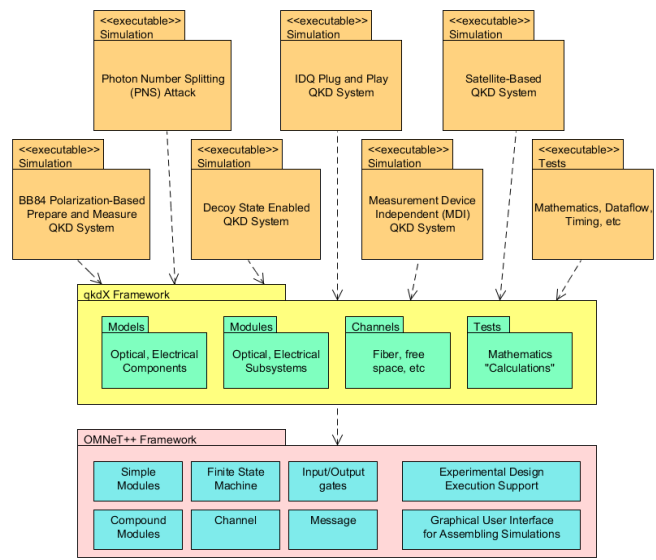

Fig. 2: Package Structure

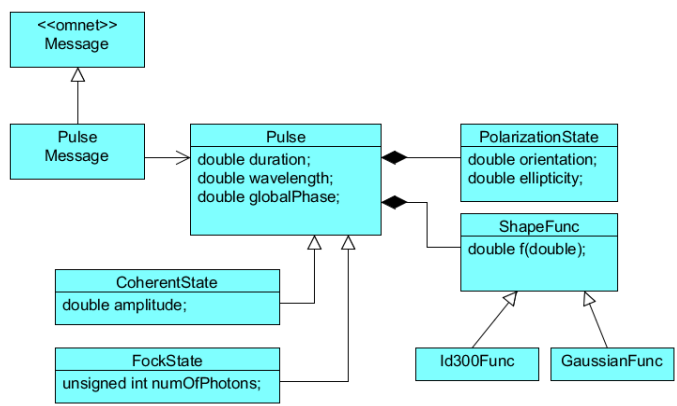

Fig. 3: Pulse Design

The shape itself is defined by C++ functor-like class, meaning shapes are defined by objects that encapsulate a single function. The abstract base ShapeFunc class defines the signature for function "f" which is overridden in subclasses. Example subclasses define simple Gaussian shapes as well as measured pulse shapes approximated by multiple Gaussians - an example of this approximation is shown for the ID Quantique ID300 [12] laser (which is used in QKD systems) in Figure 4.

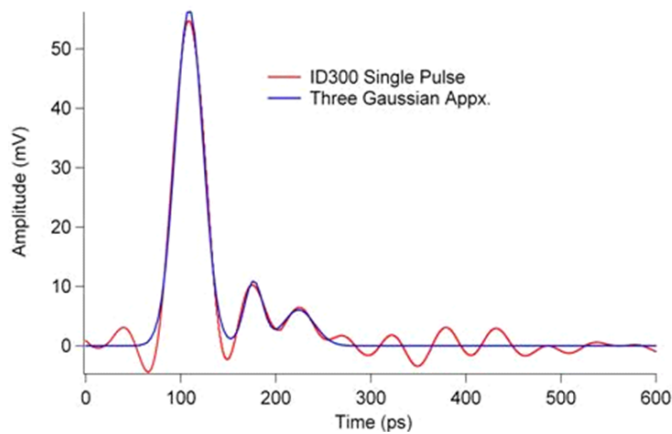

Fig. 4: ID Quantique ID300 [12] 1550nm Laser Pulse

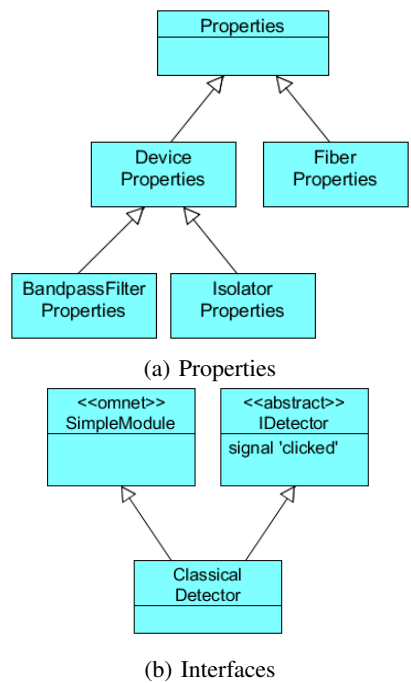

(a) Properties

(b) Interfaces

Fig. 5: Component Patterns

## V. Optical Components

Optical components, which are based on simple modules, define the behavior or unique transformations associated with pulses entering and exiting a device through specific optical ports (i.e., module gates). As an example, optical attenuators have two optical ports (i.e., gates), a circulator has three and a beamsplitter has four.

As the framework and software has matured, we have organized the code associated with modules considering four distinct perspectives: 1) as a hierarchal structuring unit with inputs and outputs (i.e., gates) which defines the flow of data (i.e., messages, pulses), 2) as the place where simulation time is known, 3) the place where mathematical calculations are performed, and finally, 4) the place where model *state* is maintained. Furthermore, we view NED file parameters from multiple perspectives as well: as inputs closely associated with the characteristics of a modeled device (i.e., the intrinsic or inherent properties that define the device), initial values for variables (i.e., *state*) that might change during simulation execution, and finally, flags (or switches) to indicate what effects should be modeled or included.

For example, a manufactured beamsplitter of a given type has some inherent properties that are listed on its' specification sheet - these properties are usually read as NED file input parameters. We explicitly differentiate these NED parameters from others, such as boolean flags to turn modeled effects on or off, or set the initial *state* for an optical device (e.g., consider the condition of device as modeled by an `enum` of the values "working," "degraded," or "damaged").

Considering these aspects, we have created a hierarchy of properties as shown in Figure 5a which is used to create "property objects" associated with each modeled optical component. These property objects separate and encapsulate the intrinsic aspects of the device of interest apart from changing state information. They are initialized during module creation by reading NED file parameters (or XML file) and used in conjunction with standalone functions to calculate modeled effects and process optical pulse transformations. Typically, the signature of the standalone functions includes a `const reference` to both its device properties, the optical pulse to be processed, and depending on purpose, *state* information.

This strategy greatly reduces the amount of code written in modules and helps reinforce their role as arbitrators or traffic cops to manage the flow of data, and a few state variables.

In a similar manner to how the INET [9] framework defines abstract interfaces (i.e., contracts), qkdX also defines interfaces for the various optical component categories (e.g., all detectors `emit` a "clicked" signal) as shown in Figure 5b.

## VI. Fiber Channels

Modeling fiber cables involves subclassing OMNeT++'s channel abstraction and adding the features needed to account for length, attenuation effects, and even polarization "walk" or "drift" properties. The same approach taken with optical components in terms of creating a property object and standalone functions to process effects was also taken with channels.

## VII. Testing

Structuring the code in terms of property objects and standalone *stateless* functions resulted in a significant amount of the codebase not resting (or subclassing) from any OMNeT++ classes. This is a natural result, given that we structured our code so that OMNeT++ modules and channels solely serve the role of traffic cops concerning data flow and the managers of simulation state data. Given this software organization, we found the entire codebase to be more testable.

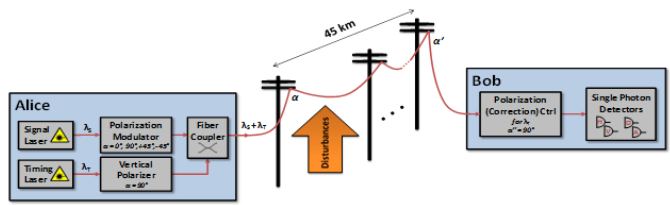

Fig. 6: Polarization Controller Performance Model

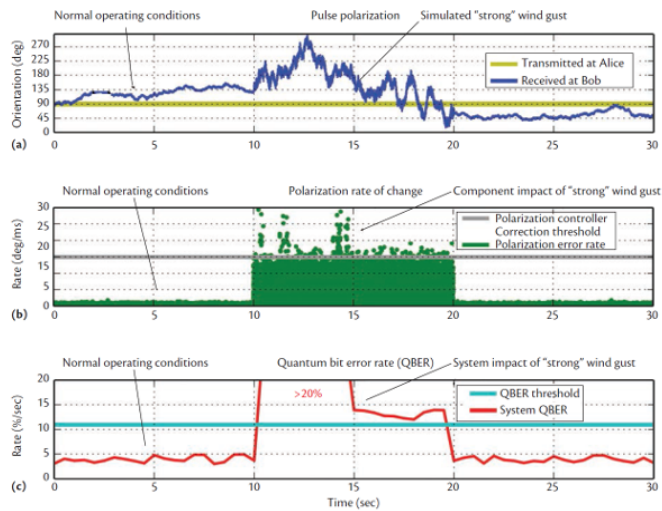

Fig. 7: Polarization Controller Performance Analysis

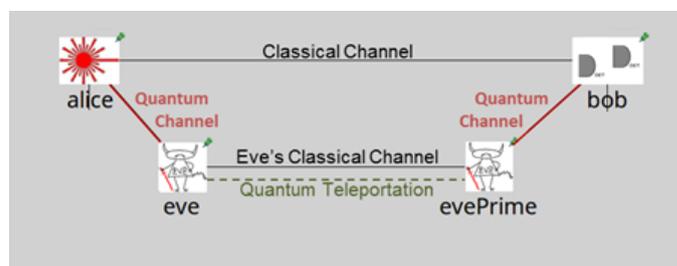

Fig. 8: Decoy State Enabled QKD System Model

Using the Simplified Wrapper and Interface Generator (SWIG) [13] utility, it became quite easy to "wrap" the C++ classes and functions with proxy interfaces so they can be imported and accessed from Python [14]. Using Python and IPython [15] in particular, testing the non-time dependent aspects of algorithms was simplified by writing relatively short scripts as compared to say, writing the same test in C++ code. Plotting results is greatly facilitated using various Python support packages such as matplotlib [16]. Another side benefit from this approach became apparent as the team testing component functions did not have to be fluent in C++ to perform this task.

For testing data flow and time dependent aspects of a model, OMNeT++-based simulations have been defined.

## VIII. System Simulation Studies

In this section, we briefly describe three different simulation studies that were conducted using the qkdX framework to demonstrate its utility.

We leveraged OMNeT++'s robust data collection and analysis tools to facilitate a deeper understanding of QKD performance-security trades with respect to relationships between quantum phenomenon (e.g., pulse propagation, temperature changes, and physical disturbances) and system-level interactions (e.g., hardware designs, software implementations, and protocols). Further, the ability to rapidly export collected data to statistical analysis programs, such as R [17], provided the ability to quickly summarize large numbers of experiments in an efficient manner.

### A. BB84 QKD Reference Architecture

Initially, we defined and modeled a polarization based, prepare and measure BB84 QKD architecture, which we refer to as the reference architecture. Because there was no published examples of QKD systems with sufficient detail necessary to implement a system-level model, we consulted with subject matter experts to define it [2]. This effort was beneficial because it forced us to formalize the behavior of each QKD functional unit.

As we documented each of the functional units, we coded OMNeT++ modules to exhibit the desired behavior. Initially, we constructed this architecture using simple and compound modules that exclusively used the handleMessage() paradigm. However, we found that for our application, in some limited cases, the activity or process-based paradigm was more readable, more understandable, and could more easily be modified.

Our first system-level study using the reference model was to examine the impact that polarization drift played in QKD system behavior. Figure 6 shows an abstraction of the system model used to study the behavior of polarization drift in aerial optical fiber in a real world physical link [4]. The simulation results confirmed experimental results that strong wind gusts in aerial fibers generate sufficient polarization drift to cause system outages. Figure 7 shows a 30-second interval during which initially the polarization is correctable, but then becomes excessive such that it could not be corrected by the polarization controller present within Bob.

### B. Decoy State Enabled QKD

As we gained more experience using the qkdX framework, we expanded our exploration into modeling processes and protocols contained in QKD systems related to operational security of the system. One such protocol is the decoy state protocol [18]. In Figure 8, we present a decoy state enabled QKD system model which extended the reference architecture to include the necessary components and behaviors required to perpetrate a Photon Number Splitting (PNS) attack against the QKD system [19]. The PNS attack is conducted by an adversarial eavesdropper, Eve, whose purpose is to gain knowledge of the generated secret key. Eve performs the PNS attack by stealing photons from each multiphoton pulse Alice generates while suppressing all single photon pulses [3]. Modeling the PNS attack required implementation of a new optical pulse representation known as the Fock state and required development of notional components. Existing optical component models were modified to handle both weak coherent pulses and Fock states. Additionally, Alice and Bob's individual processors had to be extended to perform analysis of signal and decoy qubit detection statistics to detect the presence of a PNS attack [19].

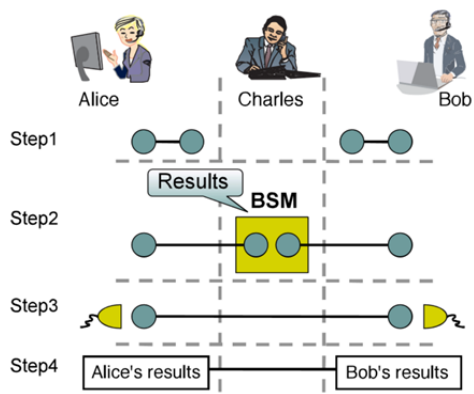

Fig. 9: MDI QKD Process [20]

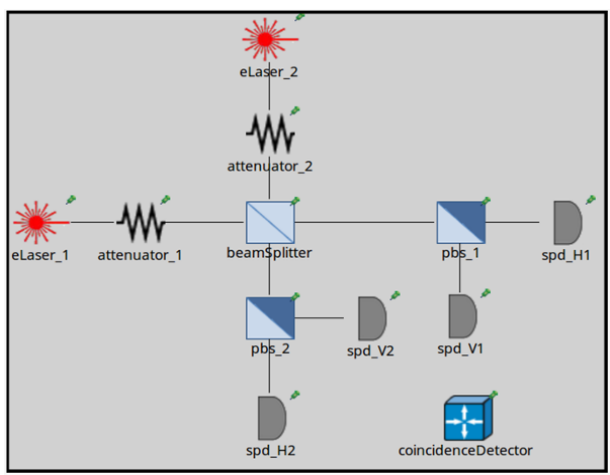

Fig. 10: MDI QKD OMNeT++ Network

*C. Measurement Device Independent QKD*

Detectors are critical components in a QKD system and their operational characteristics greatly impact the performance and security of the system as a whole. One problem identified by QKD researchers was the need to remove detector side channel attacks which can undermine the security of a QKD system. Side channels allow information to leak from the system and can arise from many sources including component imperfections, poor or malicious manufacturers, or the operational environment. This realization led to a new protocol named, Measurement Device Independent (MDI) QKD [20].

In Figure 9, we show a simplified model of a MDI QKD system. This model utilizes components of the qkdX framework to study the Bell State Measurement (BSM) [20]. MDI QKD requires Alice and Bob to individually generate entangled photon pairs. Next, Alice and Bob send half of the entangled photon pair to a third party known as Charles, who performs the BSM and reports his results to Alice and Bob. Then, Alice and Bob perform individual measurements on the retained half of the photon pair. Last, they compare a sample of their results to those reported by Charles. The comparison of these results mitigates attacks on the process by a malicious actor [20].

Figure 10 depicts the OMNeT++ network containing the components used in a MDI QKD system model. Alice and Bob are represented by each eLaser and attenuator pair, while the Bell State Analyzer (BSA) is represented with a beam splitter (BS), polarizing beam splitters (PBSs), and detectors (SPDs). This model required the extension of existing reference model laser capabilities which includes logic to control polarization and detector timing information. Additionally, the modeled beam splitters were extended to represent asymmetric behaviors. Furthermore, the SPDs were enhanced to accurately process interference from multiple pulses (i.e., the BSM) during detection periods known as gates.

Our initial results revealed some deficiencies in our original conceptualization of the MDI QKD model. Instead of attenuating weak coherent pulses down to signal photon levels, we must instead use heralded single photon Fock states to properly calculate the quantum interference effects. Work is currently underway to implement the mathematics necessary to properly account for quantum interference within a BSA apparatus.

## IX. CONCLUSIONS

In this paper, we presented the qkdX framework, a unique application of the OMNeT++ framework to the domain of modeling optical systems. The motivation to develop this framework stems from a desire to study and better understand the impact of non-idealities on QKD system performance. Using OMNeT++'s module, message and channel abstractions, models of optical components, pulses and fiber cables have been created. We have introduced our approach to modeling these systems using OMNeT++ without including too many details associated with quantum calculations.

We have also presented three QKD system models built leveraging this simulation framework. The system models have been used to efficiently study of QKD systems, protocols, and components.

## ACKNOWLEDGMENT

This work was supported by the Laboratory for Telecommunication Sciences [grant number 5743400-304-6448] and in part by a grant of computer time from the DoD High Performance Computing Modernization Program at the Air Force Research Laboratory, Wright-Patterson AFB, OH.

## Disclaimer

The views expressed in this paper are those of the authors and do not reflect the official policy or position of the United States Air Force, the Department of Defense, or the U.S. Government.

## References

[1] C. Benett and G. Brassard, "Quantum cryptography: Public key distribution and coin tossing," *Proceedings of IEEE International Conference on Computers, Systems and Signal Processing*, 1984.

[2] L. Mailloux, J. Morris, M. Grimaila, D. Hodson, D. Jacques, J. Colombi, and G. Baumgartner, "A modeling framework for studying quantum key distribution system implementation non-idealities," *IEEE Access*, 2015.

[3] G. Brassard, N. Lutkenhaus, T. Mor, and B. Sanders, "Limitations on practical quantum cryptography," *Physical Review Letters*, vol. 85, no. 6, pp. 1330–1333, 2000.

[4] L. Mailloux, M. Grimaila, D. Hodson, G. Baumgartner, and C. McLaughlin, "Performance evaluations of quantum key distribution system architectures," *IEEE Security and Privacy*, vol. 15, no. 1, pp. 30–40, 2015.

[5] "OMNeT++ Discrete Event Simulator," https://omnetpp.org, accessed: 2015-06-28.

[6] S. Wiesner, "Conjugate coding," *ACM Sigact News*, vol. 15, no. 1, pp. 78–88, 1983.

[7] B. Schumacher, "Quantum coding," *Physics Review A*, vol. 51, no. 4, pp. 2738–2747, 1995.

[8] M. Grimaila, J. Morris, and D. Hodson, "Quantum key distribution: A revolutionary security technology," *The Information System Security Association (ISSA) Journal*, pp. 20–27, 2012.

[9] "INET framework," https://inet.omnetpp.org, accessed: 2015-06-28.

[10] "Coherent states," https://en.wikipedia.org/wiki/Coherent_states, accessed: 2015-06-28.

[11] "Fock or number state," https://en.wikipedia.org/wiki/Fock_state, accessed: 2015-06-28.

[12] "ID300 series sub-nanosecond pulsed laser source datasheet," http://www.idquantique.com/images/stories/PDF/id300-laser-source/id300-specs.pdf, accessed: 2014-03-05.

[13] "SWIG simplified wrapper and interface generator," http://www.swig.org, accessed: 2015-06-28.

[14] "Python," https://www.python.org, accessed: 2015-06-28.

[15] "IPython," http://ipython.org, accessed: 2015-06-28.

[16] "matplotlib plotting package for python," http://matplotlib.org, accessed: 2015-06-28.

[17] "The R Project for Statistical Computing," http://www.r-project.org, accessed: 2015-06-24.

[18] H.-K. Lo, X. Ma, and K. Chen, "Decoy state quantum key distribution," *Physical Review Letters*, vol. 94, no. 230504, 2005.

[19] R. Engle, M. Grimaila, L. Mailloux, D. Hodson, G. Baumgartner, and C. McLaughlin, "Developing a decoy state enabled quantum key distribution model," *IEEE Transactions on Systems, Man, and Cybernetics: Systems*, 2015 submitted.

[20] F. Xu, M. Curty, B. Qi, and H.-K. Lo, "Measurement-device-independent quantum cryptography," *IEEE Journal of Selected Topics in Quantum Electronics*, 2014.